\documentstyle[graphicx]{aipproc}
\begin{document}
 
\title{Jet Induced Supernovae: Hydrodynamics and Observational Consequences}
\author{A. Khokhlov $^1$, P. H\"oflich$^2$}
\address{$^1$ Naval Research Lab, Washington DC, USA\
$^2$ Department of Astronomy, University of Texas, Austin, TX 78681, USA}

\maketitle

\begin{abstract} 
 Core collapse supernovae (SN) are the final stages of stellar evolution in 
massive stars during which the central region collapses, forms a neutron star 
(NS), and the outer layers are ejected. Recent explosion scenarios assumed that 
the ejection is due to energy deposition by neutrinos into the envelope but 
detailed models do not produce powerful explosions. There is mounting evidence 
for an asphericity in the SN which is difficult to explain within this 
picture. This evidence includes the observed high polarization, pulsar kicks, 
high velocity iron-group and intermediate-mass elements material observed in remnants, etc.
 
 The discovery of highly magnetars revived the idea that the basic mechanism 
for the ejection of the envelope is related to a highly focused MHD-jet formed 
at the NS. Our 3-D hydro simulations of the jet propagation through the star 
confirmed that the mechanism can explain the asphericities.    
 
 In this paper, detailed 3-D models for jet induced explosions of   
"classical"
core collapse supernovae are presented. We
demonstrate the influence of the jet properties
and of the underlaying progenitor 
structure on the final density and chemical structure. Finally,         
we discuss the observational
consequences, predictions and tests of this scenario.               
\end{abstract}

\section{Introduction}
 
 Supernovae (SN) are among the most spectacular events because they reach the
same brightness as an entire galaxy. This makes them good candidates to
determine extragalactic distances and to measure the basic cosmological parameters.
 Moreover, they are thought to be the major contributors to the chemical
enrichment of the interstellar matter with heavy elements.
   Energy injection by SN into the interstellar medium, triggers star formation and feedback in
galaxy formation, and is  regarded as a key for our  understanding
of the formation and evolution of galaxies.

Core collapse supernovae are thought to be the final stages of the
evolution of massive stars which live only                    
 $10^6$ to $2 \times 10^8$ years. Such supernovae could be the brightest
objects in the distant past when stars first began to form.     
  A detailed understanding of core collapse
is essential to probe the very early phases of the Universe
 right after the initial star forming period which occurs at redshifts $z \geq 3 ... 5$. 
Understanding the mechanism of core collapse supernovae explosions
is a problem that has challenged researchers for
decades (Hoyle \& Fowler 1960). 
In the general scenario for the explosion,
 the central region of a massive star    
collapses and forms a neutron star. Eventually, parts of the potential energy
will cause the ejection of the envelope.
This general scenario  has been confirmed by a wealth of observations 
including the direct  detection of neutrinos in SN1987A and neutron stars 
in young supernovae remnants. 

In recent years, there has been a mounting evidence that the explosions of massive stars (core
collapse supernovae) are highly aspherical. 
(1) The spectra of core-collapse supernovae (e.g., SN87A, SN1993J, SN1994I, SN1999em)
 are significantly polarized indicating asymmetric envelopes (M\'endez et
al. 1988; H\"oflich 1991; Jeffrey 1991; Wang et
al. 1996; Wang et al. 2000).                        
The degree of polarization  tends to vary inversely with the mass of the hydrogen
envelope, being maximum for Type Ib/c events with no hydrogen (Wang
et. al. 1999, Wang et al. 2000). For supernovae with a good time and wavelength
coverage   the orientation of the polarization vector tends to stay constant
both in time and in the wavelength. 
This suggests that there is a  global
symmetry axis in the ejecta (Wang et al. 2000).
(2) Observations of SN~1987A showed that radioactive material was brought
to the hydrogen rich layers of the ejecta very quickly during the explosion
(Lucy 1988; Tueller et al. 1991).
(3) The remnant of the Cas~A supernova shows rapidly moving oxygen-rich
matter outside the nominal boundary of the remnant             
                 and  evidence for two oppositely directed jets of
high-velocity material (Fesen \& Gunderson 1997). 
(4). Recent X-ray observations with the CHANDRA satellite have shown an unusual
distribution of iron and silicon group elements with large
scale asymmetries in Cas~A (Huges et al. 2000).
(5) After the explosion, neutron stars are observed with high
velocities, up to 1000  km/s (Strom et al. 1995).

There is a general agreement that the explosion of a massive star is caused
by the collapse of its central parts into a neutron star or, for massive 
progenitors, into a black hole.
The mechanism of the energy deposition into the envelope
is still debated.  The process likely involves the bounce and 
the formation of the prompt shock (e.g. Van Riper 1978,
 Hillebrandt 1982), radiation of the energy in the form of
neutrino (e.g. Bowers \& Wilson 1982)  and the interaction of the neutrino with the material of
the envelope and
 various types of convective motions ( e.g. Herant et al. 1994, Burrows et al. 1995,
 M\"uller \& Janka 1997, Janka \& M\"uller 1996), rotation
(e.g. LeBlanc \& Wilson 1970, Saenz \& Shapiro S.L. 1981,
M\"onchmeyer et al. 1991)
 and magnetic fields (e.g. LeBlanc \& Wilson 1970, Bisnovati-Kogan 1971).

Spherically symmetric explosion models rely on the neutrino deposition
mechanism. The results depend critically on the progenitor structure,
equation of state, neutrino physics,  and 
implementation of the neutrino transport. Currently,  results are
inconclusive even when using sophisticated Boltzman solvers for the
neutrino transport. For example, Mezzacappa et al. (2000)  find an explosion
whereas Yamada et al. (1999)  and Rammp \& Janka (2000) do not.
Even if successful, these models cannot explain
the observed asymmetries. Within the
spherical core-collapse picture, additional mechanisms must be invoked which
operate within the envelope itself.

Two such mechanisms have been studied in some detail.
 One is the Rayleigh-Taylor
instability which causes mixing of the layers of different composition when the 
 outgoing shock front passes through (M\"uller et al. 1989, Benz \& Thielemann 1990,
 Fryxell et al. 1991). This effect can explain mixing of the carbon, oxygen
and helium-rich layers required for the SN1987A, but none of the simulations
were able to account for the  high
velocity of Ni observed in SN1987A (Kifonidis et al. 2000). Rayleigh-Taylor mixing
 provides a rather small-scale structures and can hardly account for the
observed polarization which requires a global asymmetry of the expanding
envelope ( H\"oflich 1991).
Another mechanism involves an explosion inside a rapidly and differentially rotating
supernova progenitor (Steinmetz \& H\"oflich 1992). With this mechanism it was possible to account
for  the polarization in
SN1987A which originated from a blue supergiant.
This mechanism may have difficulty accounting for the early 
polarization in  some Type II supernovae (Wang et al. 2000) whose light curves
indicate a red-giant progenitors. A strong differential rotation
 in red supergiants can hardly be expected due to their convective envelopes (Steinmetz \& H\"oflich,
1991).

Attempts have also been made to include multi-dimensional effects into a
model of collapse itself.
The collapsing core becomes  unstable due to
the gradients of both electron mole fraction $Y_e$ and entropy.
The developing convection then affects both the neutrino flux and 
the energy deposition behind the stalled shock. Numerous studies
have demonstrated the presence of this effect. It is still debated
whether convection combined with the neutrino transport provides the
solution to the supernova problem (Rammp et al. 1998 and references therein).
In the current calculations, the size and scale of the convective motions 
seem to be too small to explain the observed asymmetries. The angular 
variability of
the neutrino flux caused by the convection has been invoked to
explain the neutron star kicks (Burrows et al. 1995, Janka \& M\"uller 1994). Calculations give
kick velocity up to $\simeq 100$ km/s whereas NS with velocities of several
$100$ km/s are common.

Rotation of the collapsing core may also be important. It tends to facilitate
the explosion because  the centrifugal barrier reduces the effective potential for the
material moving in the equatorial plane and introduces an axial symmetry in the fluid motions.
Simulations made so far indicate that the rotation alone has no or has only a weak effect on the
explosion (e.g. M\"onchmeyer et al. 1991, Zwerger \& M\"uller 1997).
 In the latter case it induces a rather weak asymmetry of the explosion with more
energy going along the rotational axis.

It has long been suggested that the magnetic field can play an important role in the explosion
(LeBlanc \&  Wilson 1970; Ostriker \& Gunn 1971, Bisnovati-Kogan 1971, Symbalisty 1984).
LeBlanc and Wilson 
simulations showed the amplification of the magnetic field due to rotation and the
formation of two oppositely directed,
high-density, supersonic jets of material emanating from the collapsed
core. Their simulations assumed a rather high initial magnetic field $\sim 10^{11}$ Gauss
 and
produced a very strong final fields of the order of $\sim 10^{15}$ Gauss which seemed
to be unreasonable at the time. 
The recent discovery of pulsars with very high magnetic fields 
(Kouveliotou et al. 1998, Duncan \& Thomson 1992) revives the interest in
 the role of rotating magnetized neutron stars in the
explosion mechanism. It is not clear whether a  high initial magnetic field required 
for the LeBlanc \&  Wilson mechanism is realistic. On the other hand it may not be needed.
Recently Mayer and Wilson (2000, private communication) have suggested that the field amplification and the generation of the jet
may continue during the first few 
seconds of the cooling of the neutron star when the neutron star shrinks rapidly.
The current picture of the core collapse process is unsettled. 
A quantitative model of the core collapse must eventually include all the elements mentioned above.

Due to the difficulty of modeling core collapse from first principles, 
a very different line of attack on the explosion problem has been used extensively and 
proved to be successful
in understanding of the supernova problem, SN1987A in particular
 (Arnett et al. 1990, Hillebrand \& H\"oflich 1991).
The difference of characteristic time scales of the core (a second or less)
 and the envelope (hours to days) allows one to divide the explosion problem into two largely independent
parts - the core collapse and the ejection of the envelope. 
By assuming the characteristics of the energy deposition into the envelope 
during the core collapse, the response of the
envelope can be calculated. Thus, one can study the observational
consequences of the explosion and deduce characteristics of the core collapse and the progenitor structure.
This approach has been extensively applied in the framework of the 1D spherically symmetric
formulation. The major factors influencing the outcome have been found to be
the explosion energy and the progenitor structure. The same approach can be applied in multi-dimensions to
investigate the effects of asymmetric explosions.
 In this paper we study the effects and observational consequences of an asymmetric, jet-like
deposition of energy inside the envelope of a core-collapse supernova.

\noindent
\section{Numerical Methods and Model Setup}
 
{\bf 3-D Hydrodynamics:}  The explosion and jet propagation are
 calculated by a full  3-D code within a cubic domain of size D.
The stellar material is  described by the time-dependent, compressible,
Euler equations for inviscid flow with an ideal gas equation 
with $\gamma=5/3$ plus a component due to radiation pressure with 
$\gamma=4/3$.  The Euler
equations are  integrated using an explicit, second-order accurate, Godunov type,
adaptive-mesh-refinement, massively parallel, Fully-Threaded Tree (FTT)
program, ALLA  (Khokhlov 1998).  Euler
fluxes are evaluated by solving a Riemann problem at cell interfaces.
FTT discretization of the computational domain allowes the mesh to be
dynamically refined or coarsened at the level of individual cells.
For more details, see Khokhlov (1998) and Khokhlov et al. (1999ab).

{\bf 1-D  Radiation-Hydrodynamics:}                                           
 About  1000 seconds after the core collapse and in case of the explosion of red supergiants,
the propagation of the shock front becomes almost spherical (see below). To be able to follow
the developement up to the phase of homologous expansion ($\approx 3-5$ days), the 3-D structure
is remapped on a 1-D grid, and the further evolution is  
 calculated using a one-dimensional
radiation-hydro code (e.g.  H\"oflich et al. 1998)  that solves the hydrodynamical 
equations explicitly in the comoving frame by the piecewise parabolic method 
(Colella and Woodward 1984).

{\bf Radiation Transport:} Detailed polarization and flux spectra for asymmetric explosions
are calculated using our Monte Carlo code including detailed equations of state. For details,
see H\"oflich (1995), H\"oflich et al. (1995) \& Wang et al. (1998).
 
 {\bf The Setup:}
   The  computational domain is a cube of size $L$    
with a spherical star of radius $R_{\rm star}$ and mass $M_{\rm star}$
 placed in the center.
 The innermost part with mass $M_{\rm core} \simeq 1.6 
M_{\odot}$ and radius  $R_{\rm core} = 4.5    \times  10^8$~cm, consisting
of Fe and Si, is assumed to have collapsed on a timescale much faster
than the outer, lower-density material. It is removed and replaced by a
point gravitational source with mass $ M_{\rm core}$ representing the
newly formed neutron star.  The remaining mass of the envelope $M_{env}$
is mapped onto the
computational  domain. 
 At two polar locations where the jets
are initiated at $R_{core}$, we impose an inflow with velocity $v_j$
$\rho_j$.   
 At  $R_{\rm core}$,
the jet density and pressure are the same as those of the background
material.
For the first 0.5~s, the jet velocity at $R_{\rm core}$ is kept
constant at $v_j$.
  After 0.5~s, the
velocity of the jets at $R_{\rm core}$ was gradually decreased to zero
at approximately 2~s. The total energy of the jets is $E_j$.
 These parameters are
consistent within, but somewhat less than, those of the LeBlanc-Wilson model.

\section{Results}

\subsection { Jet propagation:}
As a baseline case, we consider a jet-induced explosion in a helium
star.   Jet propagation inside the star is shown in
Fig. 1.
 As the jets move outwards, they remain collimated and do not  develop much internal
structure. A bow shock forms at the head of the jet and spreads in all
directions, roughly cylindrically around each jet.   
  The jet-engine has been switched off after about 2.5 seconds
the material of the bow shock continues to propagate through the star.                    
 The stellar material is shocked by the bow shock. Mach shocks
travels two wards the equator resulting in a redistribution of the 
energy. The opening angle of the jet depends on the ratio between the
velocity of the bow shock to the speed of sound. For a given star,    
 this angle determines the efficiency of the deposition of the
 jet energy into the stellar envelope. Here, the
efficiency of the energy deposition is about 40 \%, and
the final asymmetry of the envelope is about two.

\subsection{ Influence of the  jet properties} 
 Fig. 2 shows two examples of an explosion with
with a low and a very high jet velocity compared to the baseline case (Fig. 1).
 (Fig.1).  Fig.2  demonstrates the influence of the jet velocity on the opening
 angle of the jet and, consequently,
 on the efficiency of the energy deposition. For the low velocity jet,
the jet engine is switched off long before
the jet penetrates the stellar envelope. Almost all of the energy of the
jet goes into the stellar explosion. On a contrary, the  fast jet (61,000 km/sec)
 triggers only a weak explosion of 0.9 foe although its 
 total energy was $\approx 10 foe $.

\subsection{ Influence of the progenitor}
 For a very extended star, as  in  case of 'normal'  Type II Supernovae,
the bow shock of a low velocity jet stalls within the envelope, and
 the entire jet energy is used to trigger the ejection of the stellar envelope. In
our example (Fig. 3),
the jet material penetrates the helium core at about 100 seconds.
 After about 250 seconds the material of the jet  stalls within the
hydrogen rich envelope and
 after passing about 5 solar masses in the radial
mass scale of the spherical progenitor.
 At this time, the isobars are almost spherical,  and an almost 
spherical shock front travels outwards. Consequently, 
 Strong asphericities are 
limited to the inner regions.
 After about 385 seconds,  we stopped the 3-D run and remaped the  outer layers into 
1-D structure, and fullowed the further evolution in 1-D.
 After about 1.8E4 seconds, the shock front reaches 
the surface. After about 3 days, the envelope expands homologously.
The region where the jet material stalled, expands at velocities of about
4500 km/sec.

{\bf Fallback:}
 Jet-induced supernovae have very different characteristics with 
respect to fallback of material and the innermost structure.
 In 1-D calculations and for stars with
Main Sequence Masses of less than  20 $M_\odot$ and explosion
energies in excess of 1 foe, the fallback of material remains less than
1.E-2 to 1.E-3 $M_\odot$ and an inner, low density cavity is formed 
with an outer edge of $^{56}Ni$. For explosion energies between 1 and 2 foe,
 the outer edge of the cavity expands typically with
velocities of about 700 to 1500 km/sec
(e.g. Woosley 1997,  H\"oflich et al. 2000).
 In contrast, we find strong, continuous fallback of     
$\approx 0.2  M_\odot$ in the  the 3-D hydro models,
and no lower limit for the velocity of  the expanding material (Fig. 4).
 This significant amount of fallback  must have important consequences for the
secondary formation of a black hole.
 The exact amount and time scales for the 
final accretion on the neutron star will depend sensitively on the
rotation and momentum transport.

{\bf Chemical Structure:}
 The final chemical profiles of elements 
formed during the stellar evolution  such as He, C, O and Si
are 'butterfly- shaped' whereas the jet material fills
an inner, conic structure (Figs. 3 and 5).
 
 The
composition of the jets must reflect the composition of the innermost
 parts of the star, and should contain heavy 
and intermediate-mass elements, freshly synthesized material such
as $^{56}Ni$ and, maybe, r-process elements because, in our examples,
the entropy at the bow shock region of the jet was as high as a few hundred.
In any case,
during the explosion, the jets bring heavy and intermediate mass elements
into the outer H-rich  layers.

\section{Conclusions}

We have numerically studied the explosion of Core Collapse 
supernovae caused by supersonic jets generated in the center of the 
supernova as  a result of the core collapse into a neutron star. 
We  simulated  the process of the jet propagation through the star, 
and the redistribution of elements.  
 A  strong explosion and a high efficiency for the conversion of the
jet energy requires low jet velocities or a low, initial collimation
of the jet.  With increasing extension of the envelope, the 
conversion factor increases. Typically, we would expect higher
kinetic energies in SNe~II compared to SNe~Ib/c if a significant amount
of explosion energy is carried away by  jets. 
Within the framework of jet-induced SN, the lack of this evidence
suggests that the jets  have low velocities.

 For the compact progenitors of SNe~Ib/c, the final departures of the iso-density contours from sphericity
are typical a factor of two. This 
 will produce a linear polarization of about
2 to 3 \% (Fig. 6) consistent with the values observed for Type Ib/c supernovae.
 In case of a red supergiant, i.e. SNe~II,
the   asphericity is restricted to the inner few solar masses. In the latter case,
the iso-densities show an axis ratios  of up to $\approx$ 1.4 
at the innermost, hydrogen-rich layers.  The outer layers remain spherical.
 This has strong  consequences for the observations,  in particular,
for polarization measurements. In general, the polarization should be
larger in SNe~Ib/c compared to classical SNe~II which is consistent
with the observations by Wang et al. (2000).                    
        Early on, we expect no or little
polarization in supernovae with a massive, hydrogen rich envelope which
will increase with time to about 1 \%                            
(H\"oflich 1991), depending on the inclination  the supernovae is 
observed. This is also consistent both with the long-term time evolution 
of SN1987A (e.g. Jefferies 1991) and, in particular, the plateau
supernova 1999em which has been observed
recently with VLT and Keck (Wang et al. 2000; Leonard et al. 2000).
 
 The He, C, O and Si rich layers                     of the 
progenitor show   characteristic, butterfly-shape structures.
This  overall morphology and pattern should be observable in supernovae remnants, e.g. with the Chanda 
observatory despite some modifications and instabilities when the expanding medium interacts with the 
interstellar material.
 
   During the explosion, the jets bring heavy and intermediate mass elements
into the outer layers including $^{56}Ni$. Due to the high entropies
of the jet material close to the center, this may be a possible site
for r-process elements.
Spatial distribution of the jet material will influence the
properties of a supernova.
 In our model for a SN~II, the jet material stalled within the expanding envelope 
corresponding to a velocity of $\approx 4500 km/sec$ during
the phase of homologous expansion.
 In SN1987A, a bump in spectral lines of various elements has been
interpreted by material excited by a clump of radioactive $^{56}Ni$
(Lucy 1988). 
Within our framework, this bump may be a measure of 
region where the jet stalled.  
This could also explain the early appearance of X-rays
 in SN1987A which requires strong mixing of radioactive material
into the hydrogen-rich layers
  (see above). We note that, if this interpretation
is correct, the 'mystery spot' (Nisenson et al. 1988) would
be unrelated.
 In contrast to 1-D simulations, we find in our models strong,  
continuous fallback over an extended period of time, and a lack 
of an inner, almost empty cavity. This significant amount of fallback
and the consequences for the secondary formation of a black hole shall
be noted. Moreover, fallback and the low velocity material may alter
the  escape probablity for $\gamma $-rays produced by radioactive 
decay of $^{56}Ni$. In general, the lower escape probability is           
unimportant for the determination of the total $^{56}Ni$ production by the
late LCs   because full thermalization can be assumed in core collapse
SN during the first few years.
 However, in extreme cases such as SN98BW (e.g.
Schaefer et al. 1999), only
a small fraction of gamma's are trapped. Effects of multi-dimensionallity 
will strongly alter the  energy input by radioactive material and
disallow a reliable estimate for the total $^{56}Ni$ mass.

 Finally, we want to emphasize the limits of this study and some of the
open questions which will be addressed in future.
 We have assumed that jets are formed in the course 
of the formation of a neutron star, and have addressed observational
 consequences and constrains. However, we have not calculated the jet
formation, we do not know if they really form, and, if they form, whether
they form in all core-collapse supernovae.
We cannot claim that the jets are the only mechanism that can explain
 asphericity in supernovae although we are not aware of the others.
Qualitatively, the observational properties of
core collapse supernovae are consistent with jet-induced supernovae
and support strongly that the explosion mechanism is highly aspherical  but no
detailed comparison with an individual object has been performed.

\noindent{\sl Acknowledgments:} We want to thank our collegues for helpful discussions,
in particular, E.S. Oran, L. Wang, J.C. Wheeler,  Inzu Yi A., 
C. Mayers, J.C. Wilson, A. Chieffi,
 M. Limongi, and O. Straniero.    This work is supported in part by NASA Grant                  
LSTA-98-022.

\begin{figure}[th]
\caption {Logarithm of the density structure 
as a function of time for a helium core.
The total mass of the ejecta is 2.6 $M_\odot$.
The initial radius, velocity  and density of the jet were taken to 1200 km
32,000 km/sec and $6.5E5 g/cm^3$, respectively. The shown domains 7.9,       
9.0, 36 and 45 $\times 10^9 cm $.
The total energy is about 9E50 erg. After about 4.5 seconds, the 
jet penetrates the star. The energy deposited in the stellar envelope by the jet
is about 4E50 erg, and the final asymmetry is of the order of two.}
\caption {Same as Fig. 1 ($0.5 \leq log(\rho)\leq 5.7 $)
 but for a jet velocity of 61,000 km/sec and
a total energy of 10 foe at $\approx 1.9 sec$ (left), and 
 11,000 km/sec and a total energy of 0.6 foe (right).  The size of the
presented domains are 5 (left) and 2 $10^10 cm$ (right), respectively.
 For the high velocity jet,
 most of the energy is carried away by the
jet. Only 0.9 foe are deposited in the expanding envelope.
 In case of a low velocity jet, the bow-shock still propagates through
the star after the jet is swiched off (at $\approx 3 sec$), and
the entire jet energy is  deposited in the expanding envelope.}
\caption {Same as Fig. 1 but helium abundance (between 0 to 1)
 for the explosion of a red supergiant with
207 $R_\odot$  and 7.6 $M_\odot$. The jet velocity of 11,000 km/sec
and a total energy of 2  foe has been taken. A domain of about 
$1.4 \times 10^{12} cm $ is shown. After about 30 seconds,
the material of the bow shock penetrates the Helium core, and,        
 at about 250 seconds, the jet material 'stalls' in the hydrogen rich
layers. Subsequently a almost spherical shock front propagates through the star.
 In the final configuration, asymmetries are restricted to the layers
within the 2 to 3 solar masses of the h-rich envelope.
 After homologous expansion, the region corresponding to the stalled
shock expands with about 4000 km/sec.  
All  of the jet energy is  deposited in the expanding envelope.}
\caption {
Same as Fig. 3 but the velocity distribution in the xy- and yz plane
 for the very inner regions
at about 250 sec. Note the qualitative difference between 1-D and 
multidimensional results. In 1-D, a large, almost empty cavity is found
with expansion velocities of a about 4000 km/sec for corresponding
explosions. In multidimensional simulations shown here,
 this cavity is all but  absent. Still, after 
in multidimensional simulations material can be found up with low 
velocities. Even infall can persist over a rather extended period of time.
}

\caption {
Same as Fig. 3 but  the distribution of  O and the jet material ~~~~~~~~~~~~~~~~~~~~~~~~~~~~~~~~~~
}
\caption {
Polarization spectrum for SN1993J for an axis ratio of 1/2 for an oblate ellipsoide
in comparison with observations by
Trammell et al. (1993) are given in the left plot. On the right,
the dependence of the continuum polarization (right) and directional
dependence of the luminosity is shown 
 as a function
axis ratios for oblate  ellipsoids  seen from the equator
(from H\"oflich, 1991 \& H\"oflich et al. 1995b).
}
\end{figure}

\end{document}